\documentclass[preprint2]{emulateapj}
\usepackage{subfigure}
\usepackage{amsmath}
\usepackage{epstopdf}
\usepackage{rotating}

\slugcomment{accepted by ApJL}

\shorttitle{The pulstationally clumped RR01 stars}
\shortauthors{Kunder et al.}

\begin{document}

\title{Are the double-mode bulge RR Lyrae stars with identical period-ratios
the relic of a disrupted stellar system?}

\author{Andrea Kunder\altaffilmark{1},
Alex Tilton\altaffilmark{1},
Dylon Maertens\altaffilmark{1},
Jonathan Ogata\altaffilmark{1},
David Nataf\altaffilmark{2},
R. Michael Rich\altaffilmark{3},
Christian I. Johnson\altaffilmark{4},
Christina Gilligan\altaffilmark{5},
Brian Chaboyer\altaffilmark{5}
}
\altaffiltext{1}{Saint Martin's University, 5000 Abbey Way SE, Lacey, WA, 98503, USA}
\altaffiltext{2}{Center for Astrophysical Sciences and Department of Physics and Astronomy, The Johns Hopkins University, Baltimore, MD 21218, USA}
\altaffiltext{3}{Department of Physics and Astronomy, University of California at Los Angeles, Los Angeles, CA 90095-1562, USA}
\altaffiltext{4}{Harvard-Smithsonian Center for Astrophysics, Cambridge, MA 02138, USA}
\altaffiltext{5}{Department of Physics and Astronomy Dartmouth College, Hanover, NH 03784, USA}

\begin{abstract}
Radial velocities of fifteen double-mode bulge RR Lyrae (RR01) stars are presented, six of 
which belong to a compact group of RR01 stars in pulsation space, 
with the ratio of first-overtone period to fundamental-mode period, $P_{fo}/P_{f}\sim$0.74, 
and $P_{f}\sim$0.44.  It has been suggested
that these pulsationally clumped RR01 stars are a
relic of a disrupted dwarf galaxy or stellar cluster, as they also appear to be spatially coherent
in a vertical strip across the bulge.  However, the radial velocities of the stars presented 
here, along with proper motions from Gaia DR2, show a large range of radial velocities, proper 
motions and distances for the bulge RR01 
stars in the pulsation clump, much larger than the RR01 stars in the Sagittarius dwarf galaxy (Sgr).  
Therefore, in contrast to the kinematics of the RRL stars belonging to Sgr, and those in and surrounding 
the bulge globular cluster NGC~6441, there is no obvious kinematic signature 
within the pulsationally clumped RR01 stars.  If the pulsationally clumped RR01 stars belonged 
to the same system in the past and were accreted, their accretion in the inner Galaxy was not recent, as 
the kinematic signature of this group has been lost ($i.e.,$ these stars are now well-mixed 
within the inner Galaxy).  We show that the apparent spatial coherence reported for these stars 
could have been caused by small number statistics.
The orbits of the RR01 stars in the inner Galaxy suggest 
they are confined to the innermost $\sim$4~kpc of the Milky Way.
\end{abstract}


\section{Introduction} \label{sec:intro}

Primordial building blocks are thought to merge and form galaxy 
bulges \citep[e.g.,][]{abadi03, governato07, brooks16}, and 
large-scale photometric surveys have revealed
many filamentary substructures dubbed stellar streams \citep[$e.g.,$ see 
the "Field of Streams" from][]{belokurov06}.  Still, 
no stellar stream has been confirmed as belonging to our Milky Way (MW) bulge. 
This in contrast to the many star streams 
seen in the halo and in the disk \citep[e.g.,][]{desilva07, koposov15, antoja18}.
Many of these streams are thought to be remnants of tidally disrupted
systems such as dwarf galaxies or globular clusters, and some are thought
to originate from dynamical interactions with the Galactic bar \citep[e.g.,][]{bensby07}.

One interpretation of the absence of streams belonging to the bulge is that no significant merger 
events have occurred in the bulge since the epoch of disk formation (z$\sim$3).
This is unexpected within the context of CDM galaxy formation models, but would 
support claims that the Galaxy has undergone an unusually quiet formation 
history \citep[e.g.,][]{hammer07}.  Another interpretation is that the 
signature of a merger in the inner Galaxy just hasn't been discovered yet. 
First, the proximity of the bulge ($\sim$8~kpc) makes it unlikely that streams will be identified 
from spatial over-densities from photometry \citep[e.g.,][]{helmi99}.  Unfortunately, this is how
most of the currently known star streams have been found \citep[see review by][]{grillmair16},
and where our techniques are most familiar and refined.  New approaches and
ideas are needed for situations where photometry alone cannot lead to discoveries.
Second, the lifetimes of kinematic substructure in the inner Galaxy will be shorter
than that in the halo; an in-falling cluster/galaxy will be prone to shorter mixing timescales
in the inner Galaxy due to the more centrally concentrated mass there, resulting in
shorter relaxation times \citep[see \S7.1][]{binneytremaine}.  
This would suggest that signatures of early debris may already be washed out.
However, it has been shown that long-lived streams in the
inner Galaxy are still achievable depending on certain conditions, such as if
the stream was in resonance with the rotating bar in the inner Galaxy \citep{hattori16}.
Lastly, the inner Galaxy is a complicated mixture of inner- and outer-disk, inner- and 
outer-halo, and bulge stars \citep{robin12}, where the bulge may also have
different and overlapping stellar populations, and 
disentangling any of these populations, not to mention identifying a new
coherent substructure from the inner Galaxy, is daunting.

Still, coherent structures have been seen in the direction of the bulge.  
Perhaps the most well-known example is the Sagittarius dwarf galaxy (Sgr),
discovered as an extended group of stars with common radial 
velocities in the direction of the Galactic bulge \citep{ibata94}.  
It is our nearest confirmed galactic neighbor, and is a spectacular example
of a stellar system caught in the Galaxy's gravitational pull as being tidally torn apart.  
More recently, Ophiuchus, the innermost stellar stream known in our Galaxy, 
was identified \citep{bernard14}.  It is a thin and long ($\sim$1.6 kpc) stellar stream
which has just passed its pericenter at $\sim$3 kpc from the Galactic center
\citep{sesar15}.  Both of the these structures are relatively recent debris 
and so have likely not contributed yet
substantially to the current make-up of the Galactic bulge.

One intriguing evidence for substructure in the bulge was presented through the 
discovery of double-mode RR Lyrae stars in the inner Galaxy.  
Double-mode RRLs are stars that pulsate simultaneously in both 
the fundamental and the first-overtone radial pulsation modes, and are known
as RRd or RR01 stars.  Most often, but not always, it is the first-overtone 
period ($P_{fo}$) that has the larger amplitude than the fundamental period  ($P_{f}$).
Curiously, more than 20\% of the bulge RR01 stars form a compact group with a 
ratio of first-overtone period to fundamental-mode period of $P_{fo}/P_f\sim$0.74 
and $P_f\sim$0.44 \citep[see Figure~4,][also reproduced here in our Figure~2]{soszynski14}.  
\citet{soszynski14} therefore suggested that this group of RR01 stars 
with period ratios around 0.740 form a stream in the sky 
that may be a relic of a cluster or a dwarf galaxy tidally disrupted by the Milky Way.
Not only does the compact nature of this group suggest that these stars belong to the same 
system, but also no other RR01 stars in the Galaxy have these particular pulsation properties. 
Further, \citet{soszynski14} report that the position of these 28 ``pulsationally clumped" group of 
RR01 stars in the sky is correlated, in a sense that these stars 
cross the bulge nearly vertically with only a 1.1\% chance that they 
are drawn from the same general population as the other bulge RRL.
Therefore, instead of using photometry or kinematics to disentangle sub-structure, the
strong clumping in pulsation space point to a subset of RR01 stars in the inner Galaxy 
that appear to have originated from the same stellar structure.  Because the spatial distribution
of the pulsationally clumped RR01 stars appears to be coherent \citep{soszynski14}, 
this potential debris manifest in pulsation space should be a relatively recent addition to the MW.

Here we explore the dynamics of 15 RR01 stars toward the inner Galaxy that we
have obtained radial velocities for with the intent of investigating if the pulsationally clumped RR01 stars
are kinematically similar and moving through the Galactic bulge together.
With the Gaia DR2 data release \citep{gaiadr218},
we now have in hand proper motions of the majority of RR01 toward the inner Galaxy with
a mean precision of $\sim$5\% (note these stars are too faint for Gaia radial velocities 
or accurate parallaxes).  

\section{Data}
\subsection{Radial velocities}
We used the OGLE-III catalog of RRLs \citep{pietrukowicz12} to select RR01 stars
that fell within the footprint of the BRAVA-RR survey \citep{kunder16}.  Therefore,
the spectra was taken using the AAOmega multi-fiber spectrograph on 
the Anglo-Australian Telescope (AAT) covering the wavelength 
regime of about 8300\AA~to 8800\AA~at a resolution of R$\sim$10,000.  
The observations were taken on four different observing runs:  (1) field ($l$,$b$)=(3,$-$3) observed on 2014 Jun 21 -- NOAO PropID: 2014A-0143,
(2) field ($l$,$b$)=(3,$-$5)  observed on 2015 Aug 19-20  -- NOAO PropID: 2015B-071, 
(3) field ($l$,$b$)=(6,$-$5) observed on 2016 Aug 09-10 -- 2016B-0058, 
(4) field ($l$,$b$)=(6,$-$5) observed again on 2017 Jun 18 -- 2017A-0195.
Exposure times were between one to two hours, and in general there are 
1 - 4 epochs for each RR01 star.  The reductions were carried out in conjunction with the
BRAVA-RR reductions \citep{kunder16}.

\begin{figure*}
\centering
\mbox{\subfigure{\includegraphics[height=6.0cm]{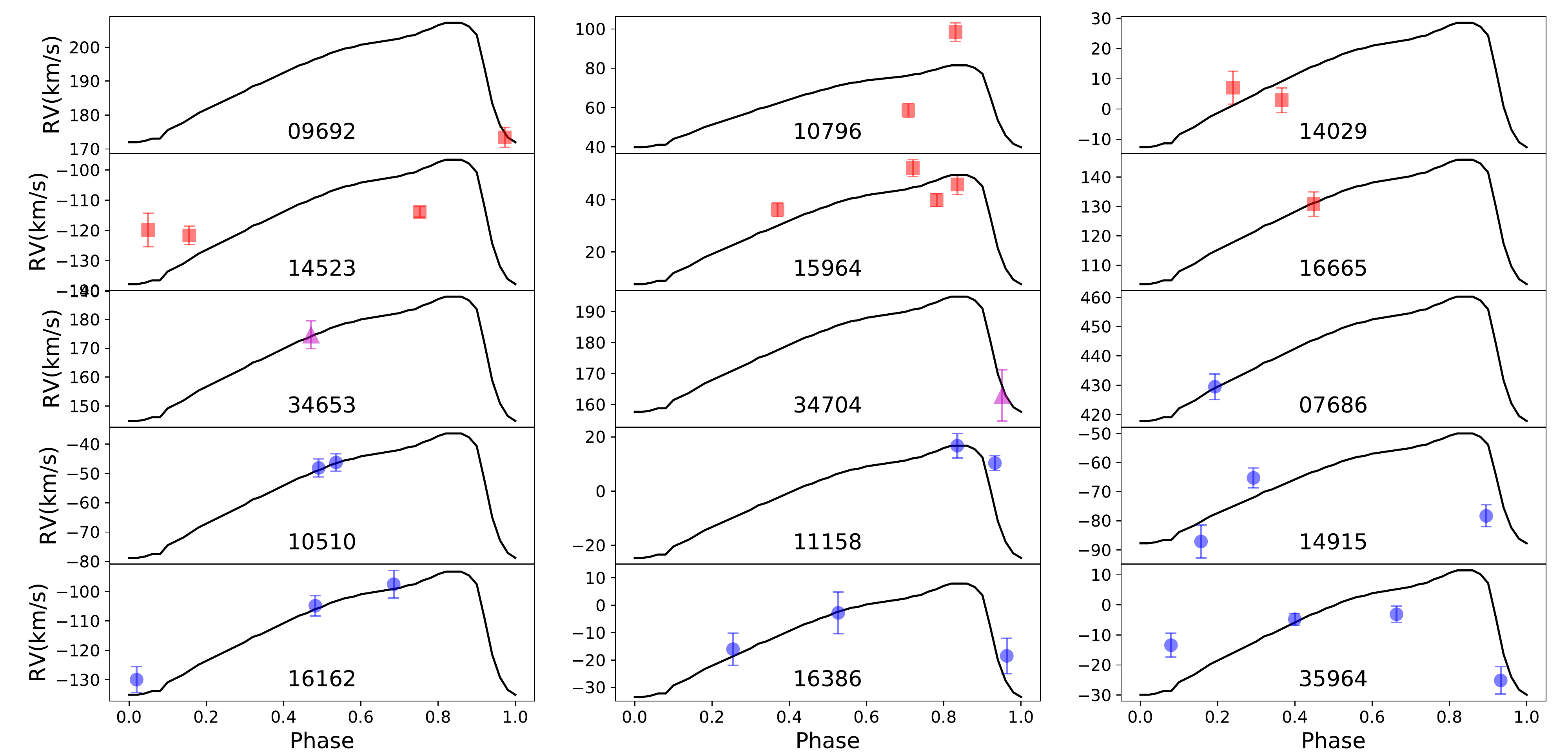}}
          \subfigure{\includegraphics[width=6.0cm]{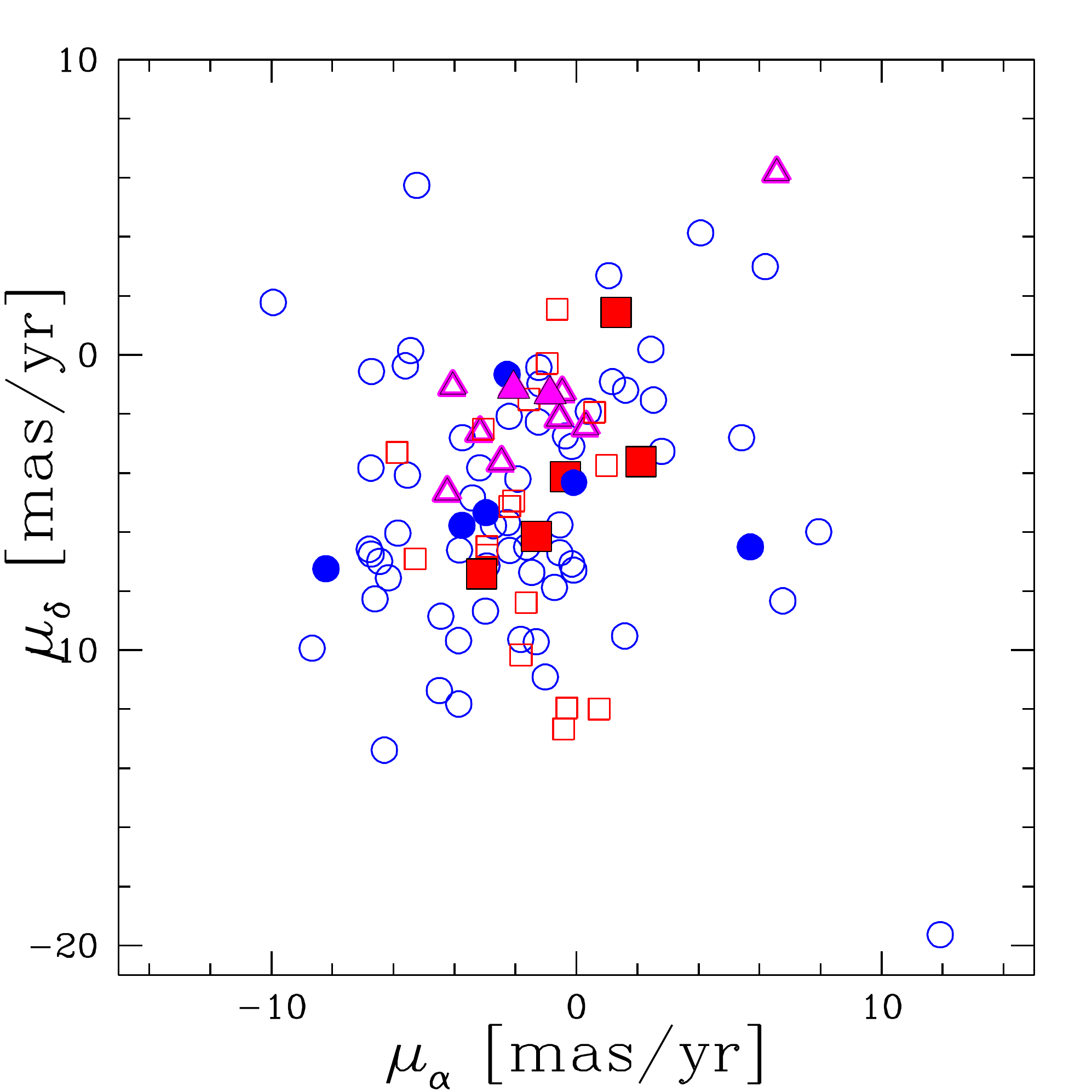}}}
\caption{{\it Left:} Radial velocity curves of observed double-mode RR Lyrae stars. The OGLE-ID for the 
RR01 stars is also shown. Stars from the grouping in $P_{fo}/P_{f}$ space are indicated by red squares. 
Stars from the Sagittarius dwarf galaxy are indicated by magenta triangles. 
Stars from the galactic bulge not in the $P_{fo}/P_{f}$ grouping are indicated by blue circles.
{\it Right:} The Gaia DR2 proper motions of double-mode RRLs.  Stars for which we have 
calculated radial velocities are shown as filled symbols, and those without radial velocity information are 
left open.  Note that the Sgr RR01s cluster 
in proper motion space.
}
\label{rvcurve}
\end{figure*}

The radial velocity curves are shown in Figure~\ref{rvcurve} and presented in Table~\ref{lcpars}.
Table~\ref{lcpars} gives the OGLE-ID (1), the RA (2) and Dec (3) as provided by OGLE, 
the star's time-average velocity (4), the number of epochs used for the star's time-average velocity (5), 
the fundamental period of the star (6), the first-overtone period of the star (7), $V$-band magnitude (8),
the $I$-band magnitude (9) and the $I$-band amplitude (10) as calculated by OGLE, 
and lastly the distance adopted for the orbital integration (11).

The RR01 radial velocity measurements have been phased by the stars known period, and 
over-plotted with the radial velocity template 
from \citet{liu91}.  This template is designed for fundamental mode 
RRL (RR0 pulsators), and the shape is slightly different RR01 stars (as discussed below).  
The template is scaled using a correlation between the amplitudes of velocity curves and light curve:
\begin{equation}
A_{rv} = \frac{40.5\times V_{amp} + 42.7}{1.37}
\end{equation}
as shown in \citet{liu91}.  As in \citet{sesar12} and \citet{kunder16} we adopt
$p$= 1.37, the so-called ``projection factor", to relate our observed radial velocities to
the pulsation radial velocities.
OGLE samples the $I$-band much more frequently than the $V$-band, so we take the
OGLE $I$-amplitude and multiply it by 1.6 to obtain each stars 
$V$-amplitude \citep[see e.g., Table~3 in][]{kunder13}.

The main uncertainties in our radial velocity measurements arise from the following factors:
(1) We have 1-4 epochs of observations, so not enough to trace out the full RR01 velocity curve.
(2) To compensate for the small number of measurements, we use RR0 light curve templates to
find the center-of-mass velocity, as no RR01 light curve templates exist in the literature.
(3) Our signal-to-noise (SNR) ratio is low, with SNR $\sim$5-20, so our individual radial velocity 
measurements have uncertainties of $\sim$5 km~s$^{-1}$.  
To understand how the first two issues affect our radial velocity measurements, we used  
the RR01 velocity curves in the globular cluster M3 \citep[][see their Figure~2]{jurcsik17}.  We 
reduced their $\sim$100 individual measurements spread out over the full pulsation cycle to
only three measurements.  
We found that the typical error between the \citet{jurcsik17} measurements using the entirety of the 
data and the measurements using the decreased data 
is $\sim$5 - 15~km~s$^{-1}$, depending on what randomly kept three measurements were retained.  
In particular, stars with observations only on the rising branch ($i.e.$, those between a phase of
$\sim$0.8 and 1.05) are the most susceptible to large uncertainties.
We therefore adopt our radial velocity uncertainty as $\sim$15~km~s$^{-1}$, 
which also encompasses the uncertainty in finding individual radial velocity 
measurements.  We note that the stars 09692, 10796 and 34704 likely have the largest 
radial velocity uncertainties, as our observations for these stars fell on the rising branch 
of their radial velocity curves.   

\subsection{Proper motions}
Proper motions of the RR01s are obtained from cross-matching with the Gaia DR2 catalog 
(Gaia Collaboration et al. 2018).  We find that 145 of the 173 RR01 stars have a proper motion 
in the Gaia DR2 catalog, where it is mainly the fainter (e.g., the Sgr RR01 stars) that do not 
have proper motions. 

\subsection{Distances}
To obtain distances to the RR01 stars, first reddenings along the line of sight of each star 
is obtained from the OGLE bulge extinction maps in \citet{nataf13}.  This extinction map was 
established using the OGLE-III passbands, similar to the OGLE-IV photometry used here, 
and makes use of the $\rm E(J-K_s)$ reddening from \citet{gonzalez12} to allow 
the coefficient of selective extinction, $\rm R_{JKVI}$ to vary, which is more realistic in the bulge region.

\begin{figure*}
\centering
\mbox{\subfigure{\includegraphics[width=3.5in]{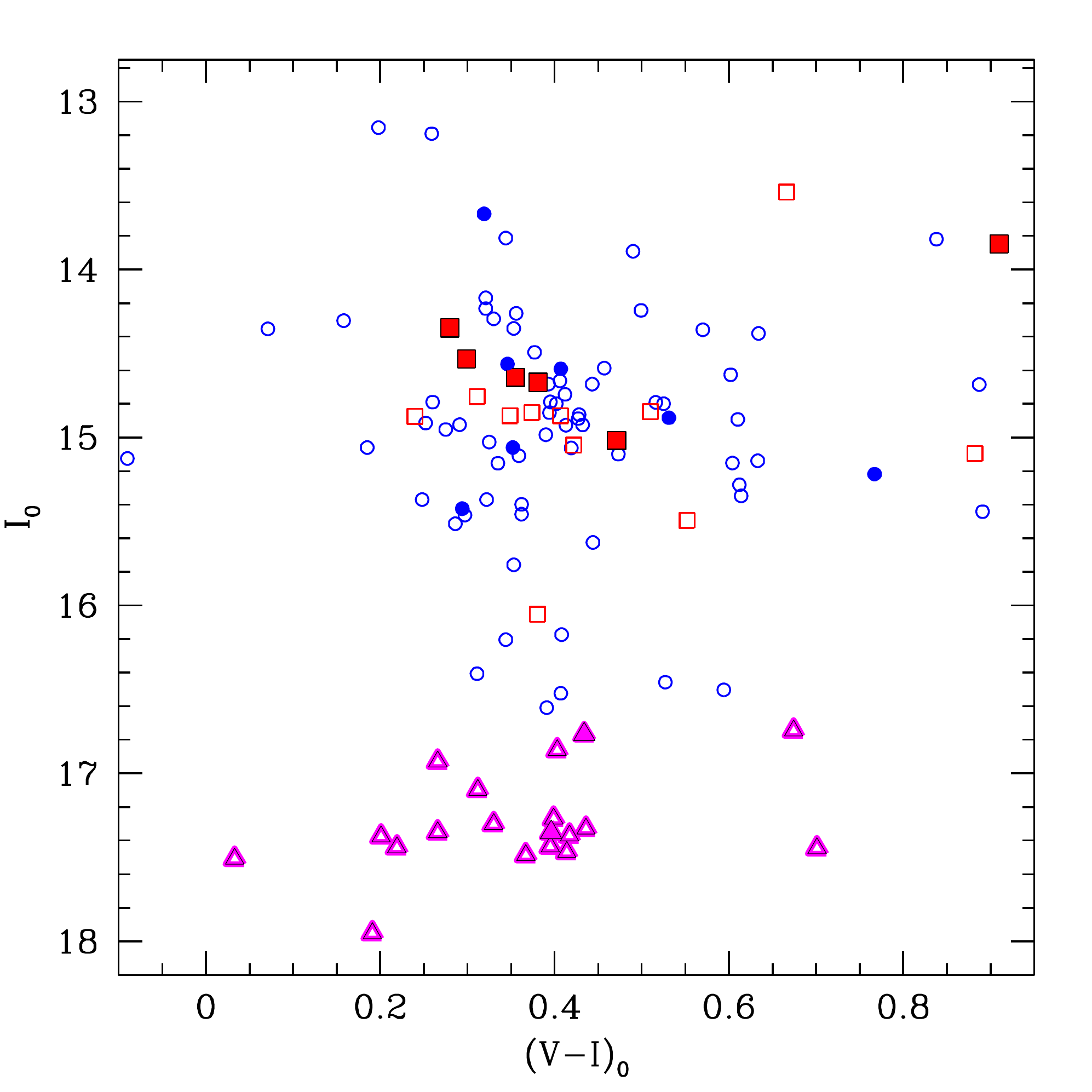}}
          \subfigure{\includegraphics[width=3.5in]{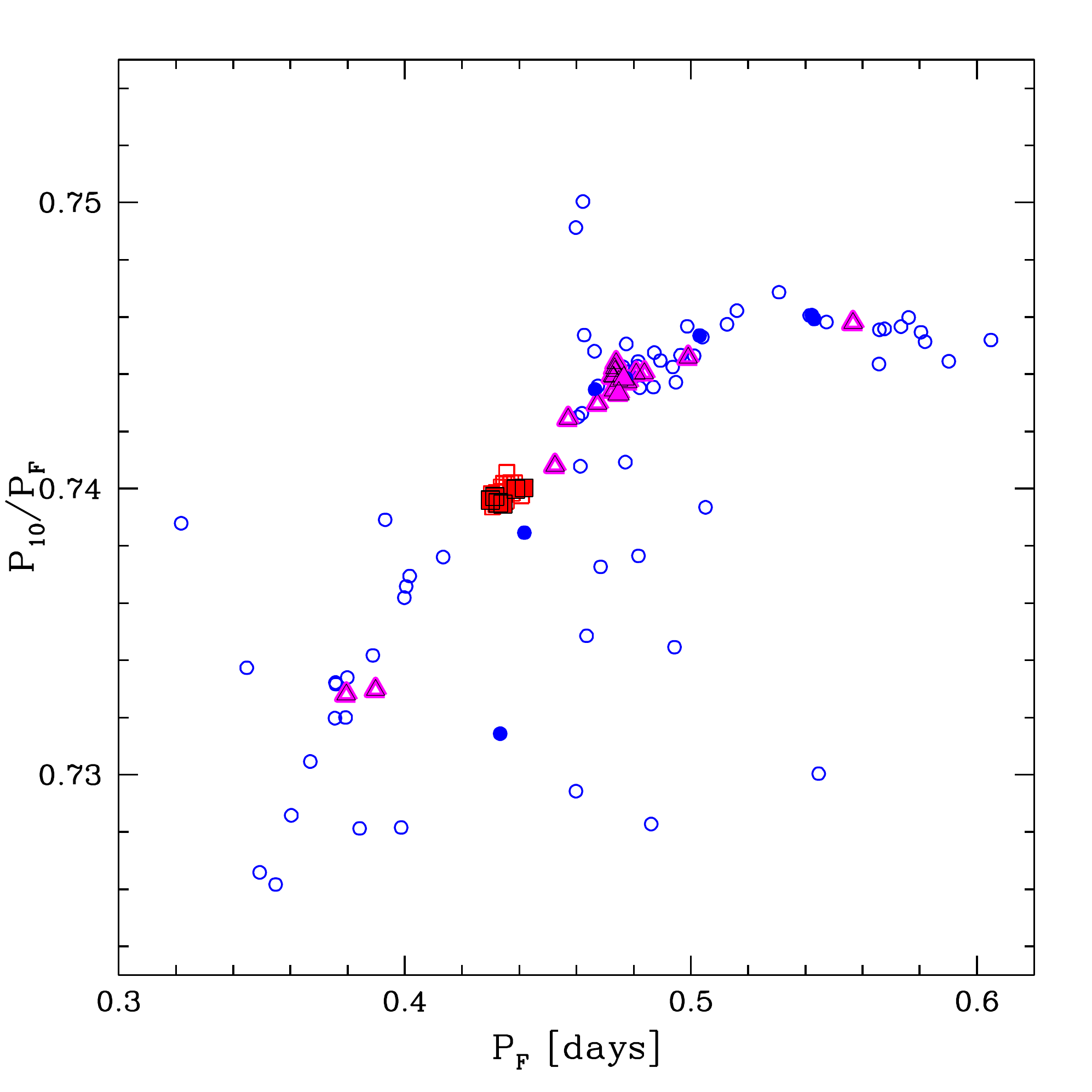}}}
\caption{{\it {Left:}}. A dereddened color-magnitude diagram of the double-mode RR Lyrae stars for those
stars with reddening information from \citet{nataf13}. 
 {\it Right:}  The period properties of these bulge RR01 stars.  
Symbols are as defined in the caption of Figure~\ref{rvcurve}.  
}
\label{deredcmd}
\end{figure*}

The dereddened color-magnitude
diagram is shown in Figure~\ref{deredcmd}.  
The stars with $I_0$ greater than $\sim$16.7~mag belong to the Sagittarius dwarf galaxy (Sgr). 
Stars with $I_0$ $\sim$16 to $\sim$16.7 may also be Sgr stars, however, their proper motions and 
$P_{fo}/P_{f}$ ratio are more consistent with RRLs belonging to the inner Galaxy.

To convert dereddened magnitudes to distances, we use the
procedure outlined in \citet{alcock97} to calculate luminosities of the RR01 stars.
This is based on pulsation equations for model envelopes of RRLs 
from \citet{bono96} and also the assumption that there is a close similarity between 
the temperature of the blue edge (FBE) of the fundamental instability strip and the 
transition zone occupied by the multimode stars 
\citep{bono94}.  Therefore, the period of the fundamental mode at the blue 
edge of the instability strip can be directly related to a stars 
temperature \citep{sandage93a, sandage93b} and hence luminosity with the equation
\begin{equation}
\log(L/L_{\sun}) = 2.506 + 2.405 \log P.
\end{equation}
This equation yields an optical RR01 distance modulus of 
18.48 $\pm$0.19~mag\citep{alcock97} to the LMC, which is one
that is consistent with the distance found from eclipsing binaries \citep{pietrzynski13}.
At $\rm P_0 =$0.46 days, the approximate period of the bulge RR01 stars in the pulsation 
clump, $\rm log L/L_{\sun}$=1.69.  Using an adopted bolometric correction of 
0.09 mag \citep{vandenberg85}, this translates to $M_V =$+0.49 for stars at $\rm P_0 =$0.46 days. 
This is well within the range of what is found for luminosities of bulge RRLs from other 
studies \citep[e.g.,][]{lee16}, as well as what is found for local RRLs
with metallicities similar to that of the bulge ($\rm [Fe/H]\sim$-1.0~dex) derived 
from a Baade-Wesselink analysis \citep{kovacs03, bono03}.

\begin{figure}
\centering
\mbox{\subfigure{\includegraphics[width=3in]{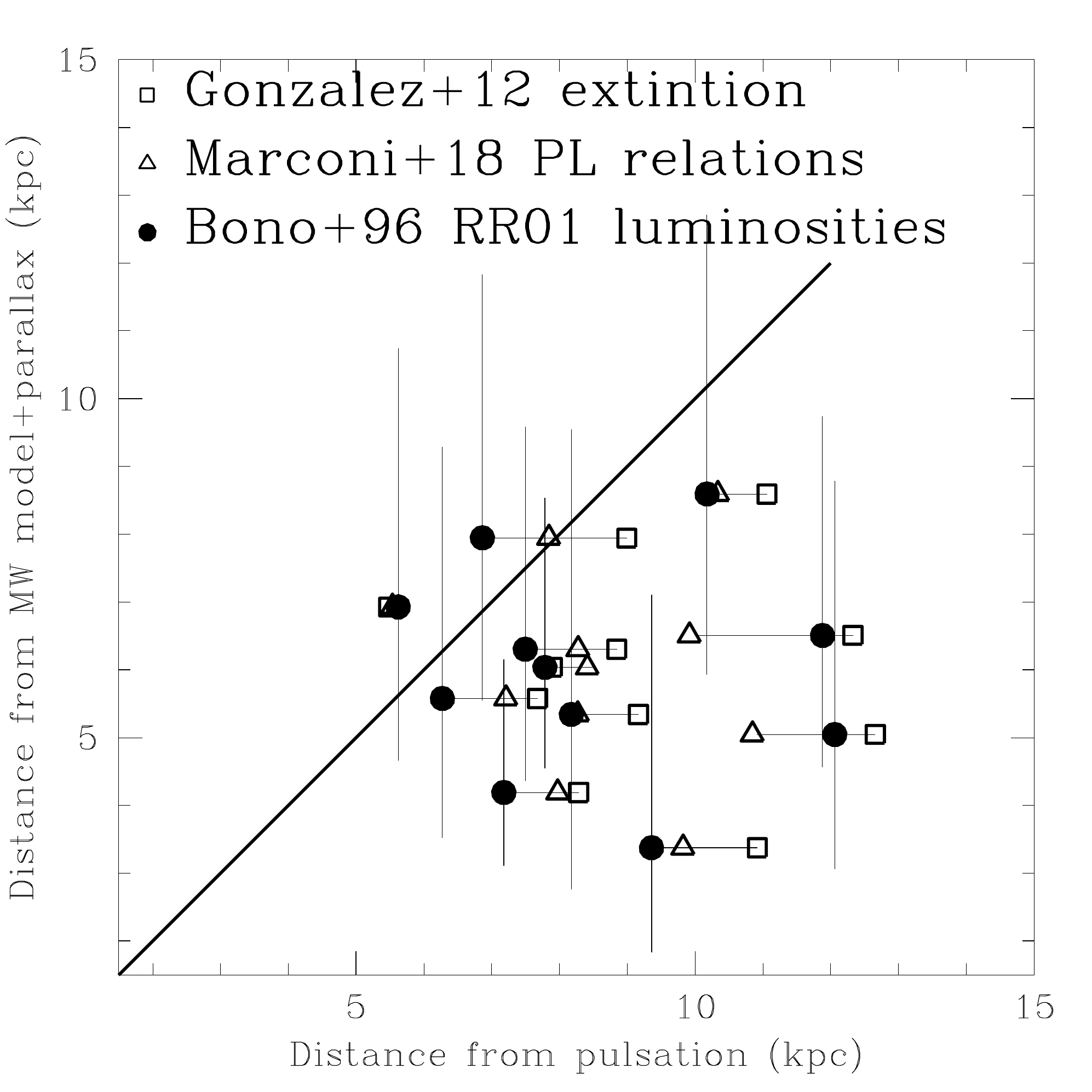}}}
\caption{The distances of our observed double-mode RR Lyrae stars obtained from their 
pulsation properties as compared to distances found using DR2 parallaxes with 
a weak distance prior that varies smoothly as a function of Galactic longitude 
and latitude according to a Galaxy model \citep{bailerjones18}.  The 
closed circles indicate distances found using pulsation equations for RR01
stars from \citet{bono96} and reddening values from \citet{nataf13}.  
The open triangles indicate distances obtained in the same manner, but
with reddening values from \citet{gonzalez12} and using the \citet{fitzpatrick99}
extinction curve.
Open squares indicate distances found using 
new Period-Luminosity-Metallicity-Helium (PLZY) relations from 
\citet{marconi18} and reddening values from \citet{nataf13}.
Different pulsation equations and reddening values lead to distance 
uncertainties of $\sim$1~kpc.
}
\label{distcomp}
\end{figure}

Figure~\ref{distcomp} shows how our distances would change if we use the \citet{gonzalez12}
reddenings and assume an extinction curve from \citet{fitzpatrick99}.

We also obtained distances following a similar procedure used for bulge RR0 stars
in \citet{pietrukowicz15}, but using newer
theoretical absolute magnitude relations for the RRLs \citep{marconi18}.  Briefly, 
we first use the mean-flux magnitudes as listed by OGLE-IV. 
We then find the absolute magnitudes $M_V$ and $M_I$ from
the theoretical relations of \citet{marconi18}:
\begin{equation}
M_V = 0.22 \rm{[Fe/H]} - 2.94 \log Y - 1.08
\end{equation}
and
\begin{equation}
M_I =0.471-1.132 \log P_f +0.205 \log Z,
\end{equation}
where $P$ is the pulsation period as determined by OGLE, 
log $Z$ = $\rm [Fe/H] -$ 1.785, and $Y$ is the helium abundance,
which we take as $Y=$0.245 \citep{marconiminniti18}.
For $\rm [Fe/H]$ we use $-$1.0~dex \citep{walker91}.
For the pulsation period, because the first overtone period is more accurate than the
fundamental mode one, we ``fundamentalized" the first overtone 
period \citep[e.g.,][]{vanalbada73, bono97, marconi03}
using
\begin{equation}
\log P_f = \log P_{fo} + 0.127.
\end{equation}
The distance can then be found using $A_I$ values from \citet{nataf13} and
\begin{equation}
d = 10^{0.2({I_0-M_I+5} )} pc. 
\end{equation}

Our distances can be compared with those found using DR2 parallaxes with 
a weak distance prior that varies smoothly as a function of Galactic longitude 
and latitude according to a Galaxy model \citep{bailerjones18}.  All our distances fell 
within the possible values allowed by the \citet{bailerjones18} error 
margins with the exception of three stars, 14523, 11158 and 16162.  
The distances from \citet{bailerjones18}
assumes the RRLs peak at a distance of $\sim$6~kpc instead of $\sim$8~kpc, 
which is the distance of the bulge \citep{blandhawthorn16}.  This shorter distance is unlikely, as these
are RRLs in the direction of the bulge with a magnitude distribution consistent with 
them being located in the inner Galaxy (see Figure~\ref{deredcmd}).  In contrast, 
the distances using RR01 pulsation properties peak at $\sim$8~kpc, in agreement 
with other distances of the RRL population toward the inner Galaxy \citep[e.g.,][]{dekany13}.

\subsection{Orbital information}
Combining right ascension ($\alpha$) and
declination ($\delta$) positions, distances, proper motions in $\alpha$ and $\delta$, 
and radial velocities, Galacto-centric spherical polar components of the velocities (radial $v_r$, 
azimuthal $v_\theta$) for our RR01s were calculated. 
We adopted a left-handed Galactic Cartesian coordinates, so that the x-axis
(i.e., U velocity) is positive going away from the 
Galactic center, the y-axis (i.e., V velocity) is positive in the direction of Galactic rotation, and 
the z-axis (i.e., W velocity) is positive towards 
the North Galactic Pole (NGP).  The local standard of rest is $\rm v_{LSR} = 220~km~s^{-1}$
\citep[e.g.,][]{bovy12},
and the distance to the Galactic center adopted is 8.2~kpc \citep{blandhawthorn16}.
Orbital information is calculated using the {\tt galpy} Python 
package\footnote{http://github.com/jobovy/galpy; version 1.2}, where 
the potential adopted is the recommended Milky-Way-like 
potential {\tt MWPotential2014} \citep{bovy15}.  

\section{Discussion}
Unlike the RR0 stars (fundamental-mode RRL), the RR01 variables are rare in the bulge 
of our Galaxy \citep[this is especially evident in the OGLE-IV catalog presented in][]{soszynski14}.  
This is also the case for RR01 stars in Galactic globular clusters and in the 
field \citep[e.g.,][]{clement01, szczygiel07}. 
They are instead more frequent in the field of other Local Group dwarf galaxies, 
where new surveys using wide-field detectors and continuous monitoring are constantly 
increasing their number\citep[e.g.,][]{kovacs01, dallora06, kinemuchi08, coppola15}. 
For example, only $\sim$0.5\% of the OGLE bulge RRL are RR01 stars \citep{soszynski14}, whereas 
there is a $\sim$5\% incidence rate of RR01 stars in the OGLE 
LMC sample, and more than 10\% of RRL are double-mode pulsators in the OGLE 
SMC sample \citep{soszynski16}. 

With exception of the bulge, all systems with double mode pulsators 
are metal-poor (with $\rm [Fe/H] < -$1.5~dex), and all show a ranking in a sense
that the $P_{fo}/P_f$ ratio becomes smaller in more metal-rich systems.  
For the bulge, the small RR01 period ratios indicate an extended metal-rich component which 
no other region in the Galaxy (or dwarf galaxy) harbors.  Hence, these objects offer a 
unique way to trace properties of an old regime that is not possible in the halo.  \citep[Note that 
the metal-rich globular clusters in the Galaxy in which RRL reside, NGC~6441 and NGC~6338, 
do not have RR01 stars; see Catalogue of Variable Stars in Galactic Globular Clusters in][]{clement01}.

Masses of RR01 pulsators are evaluated from the ratio between the first-overtone 
($P_{fo}$) and the fundamental ($P_f$) pulsation periods and pulsation models 
trace loci of constant mass in a diagram that plots the $P_{fo}/P_f$ ratio versus $P_f$, 
from which stellar masses can be estimated \citep{bono96}.  Therefore, they provide an estimate 
of the mass and the mass-metallicity relation of horizontal branch stars.
The similarity in $P_{fo}/P_f$ and $P_f$ of the pulsationally clumped RR01s indicate they have little spread 
in masses and chemistry, further suggestive that these stars belong to a unique system.

\begin{figure*}
\centering
\mbox{\subfigure{\includegraphics[width=9cm]{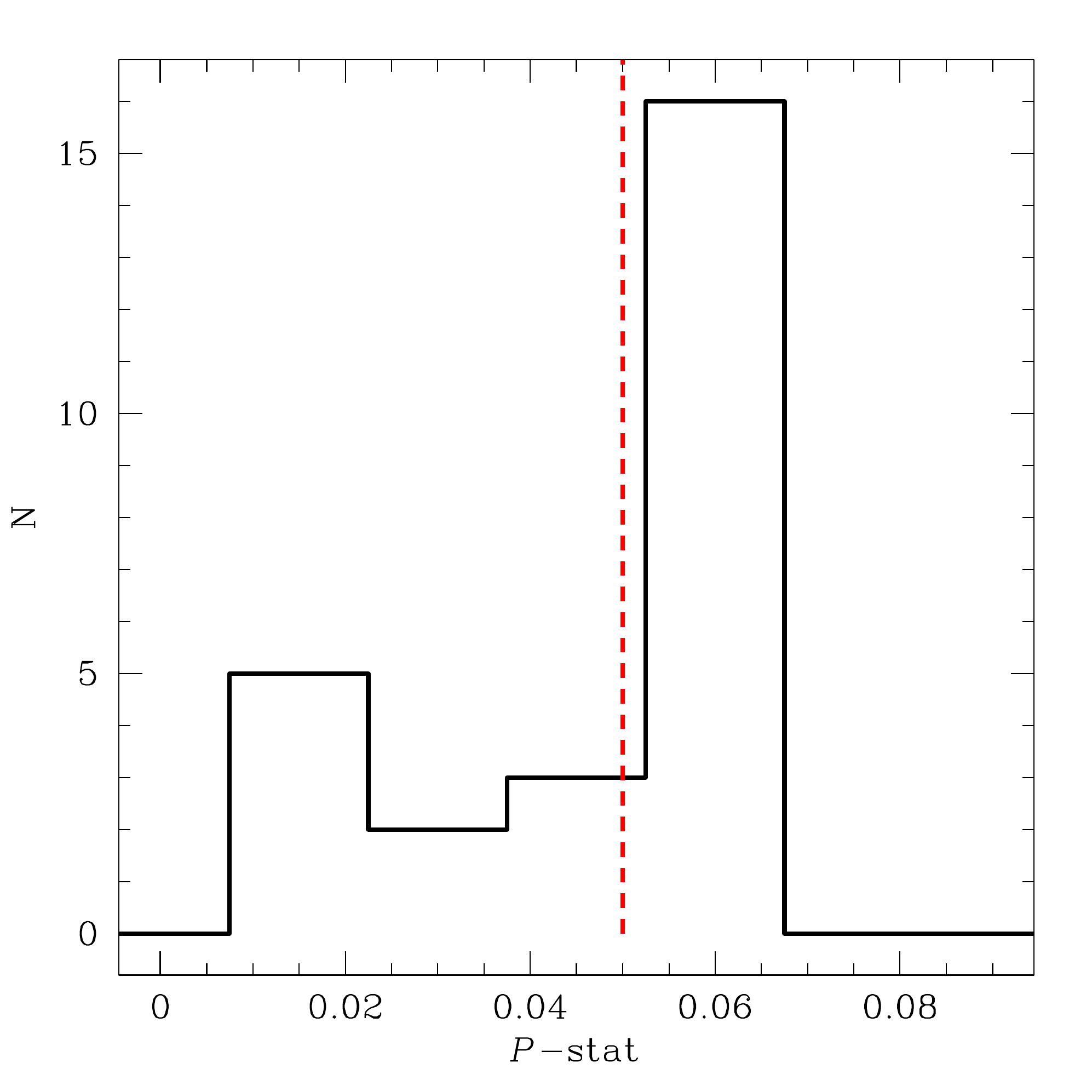}}
          \subfigure{\includegraphics[width=9cm]{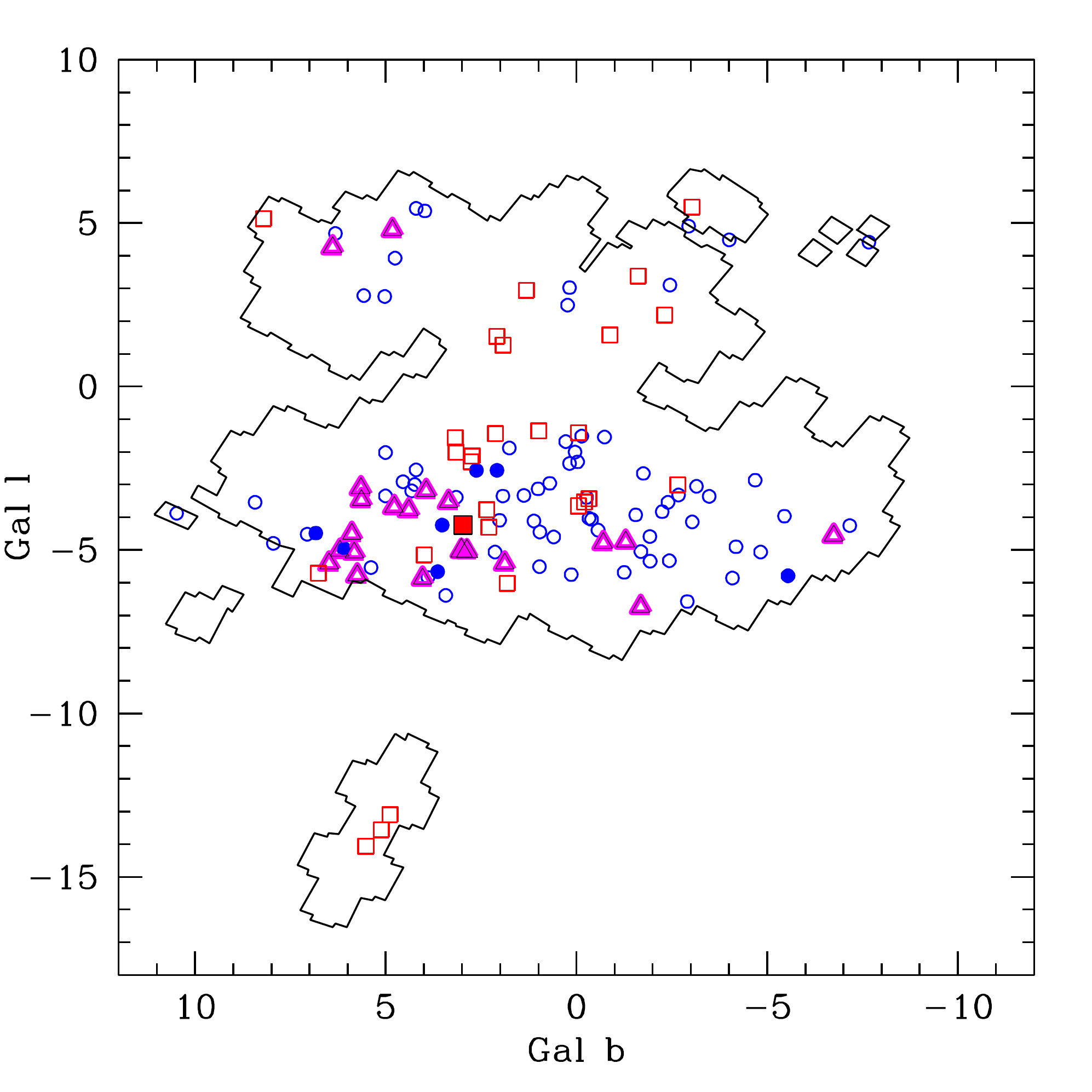}}}
\caption{
{\it Left:} We ran a 2-dimensional K-S test to establish whether one can reject the null 
hypothesis that the pulsationally clumped sample of RR01 stars and the regular 
RR01 stars come from the same distribution in Galactic latitude and longitude space. 
This histogram shows the spread in $P$-statistic values from the K-S test if 
one pulsationally clumped RR01 star is removed from the 
sample. The small sample of pulsationally clumped RR01 stars make it difficult 
to discern if the $P$-statistic is above the default threshold $\rm P_{th} =$ 0.05 below 
which one rejects the null hypothesis.  
 {\it Right:}  The spatial location in Galactic coordinates (in degrees) of the double-mode 
 RRLs shown in Figure~2; symbols are also defined there.  
 The Galactic center is at ($l$,$b$)=(0,0), and the closed
 regions show the OGLE-IV search areas. 
}
\label{pstat}
\end{figure*}

Radial velocities of 15 double mode RR Lyrae stars are calculated and presented in Table~\ref{lcpars} 
and Figure~\ref{rvcurve}.  Six of these belong to the clump in $P_{fo}/P_f$ period space.  The 
radial velocities of these 6 RR01 pulsationally clumped stars span a wide range of values, covering a 
range of $\sim$300~km/s.  
Two groups to two RR01 stars are spatially closer to each other than the other RR01s in our sample.
First, 10796 and 10510, are 0.72 degrees apart, with ($l$,$b$)=(2.7607, $-$2.3070), and ($l$,$b$)=(2.0830, $-$2.5628), respectively.  
They have radial velocities, however, that span a large range, with 63 km~s$^{-1}$ and $-$51 km~s$^{-1}$, respectively.
Second, 14029 and 14523, are 0.69 degrees apart, with ($l$,$b$)=(2.2868, $-$4.3085), and ($l$,$b$)=(2.9757, $-$4.2343), respectively.  
They also have radial velocities that span a large range, with 14 km~s$^{-1}$ and $-$115 km~s$^{-1}$, respectively.
Although disrupted clusters should show some radial velocity dispersion, especially if spread over
degrees over the sky, such a large radial velocity spread between stars within $\sim$1$^\circ$ spatial 
proximity is not expected nor seen from other 
recent accretion events \citep[e.g.,][]{anguiano16, huang19}.

Further, the proper motions of all pulsationally clumped RR01 stars 
span almost the full range of proper motion space, and there is clearly no correlation in the 
proper motions of the stars with similar period-ratios $P_{fo}/P_{f}$ (Figure~\ref{rvcurve}).  
This is in contrast to the Sgr stars, that have 
similar proper motions and radial velocities, as well as a clumping of RR01s with period 
ratios of $P_{fo}/P_{f}$ = 0.744 and $\rm P_{f} = $0.47 d (see Figure~\ref{deredcmd}).

All three components of velocity for the pulsationally clumped RR01s indicate they are 
not a coherent moving group.
The spatial coherence observed by \citet{soszynski14} may be due to
small number statistics, 
as the sample size of the pulsationally clumped RR01 stars in OGLE-IV is 28. 
The Kolmogorov-Smirnov (K-S) goodness-of-fit test used by \citet{soszynski14} 
requires a sufficient sample size, where n$>$50 is a good rule of thumb \citep[e.g.,][]{fasano87}.
We also have performed a 2-dimensional K-S test to establish whether one can reject the null 
hypothesis that the two samples of RR01 Galactic latitudes and longitudes (pulsationally clumped RR01s 
and regular bulge RR01s) come from the same distribution. 
the K$-$S test returns a probability of $P$ = 0.038, below the default threshold $\rm P_{th} =$ 0.05 below 
which one rejects the null hypothesis.  On this basis, as the OGLE team first reported, it could be assumed 
that the spatial distribution of the pulsationally clumped RR01s and the regular bulge RR01s are not drawn from 
the same sample.  

To test how the small sample size affects the K-S test, we removed one star from the 
sample of 28 pulsationally clumped RR01 stars and re-calculated the $P$-statistic.  The removed 
star was then re-inserted in the sample, and the next star was removed in an iterative manner.
Figure~\ref{pstat} shows how the $P$-statistic varies if one pulsationally clumped 
RR01 star is removed from the sample.  We see that the $P$-statistic does 
have a noticeable effect on the exact value of the $P$-statistic of a KS-test when using this
small sample of stars.  Therefore, it is unclear from the KS-test alone 
if the pulsationally clumped RR01s are a spatially coherent structure extending across the bulge.

Also, although the main body of the bulge is observed relatively symmetrical, 
the OGLE fields extend to $b$$\sim$15$^\circ$ on only one side of the bulge, so as to 
observe the main body of the Sgr Galaxy.  Three pulsationally clumped RR01 stars are found in the main 
body of Sgr, which just happens to be located along the proposed 
nearly vertical direction of the RR01 ``stream".  However, no OGLE fields are observed on the 
other side of the bulge.  Therefore, this vertical ``alignment" may be due to the fact that RR01 stars
were not able to be discovered in the other directions compared to the main body of the bulge.

\begin{figure*}
\centering
\subfigure{
\includegraphics[height=6.0cm]{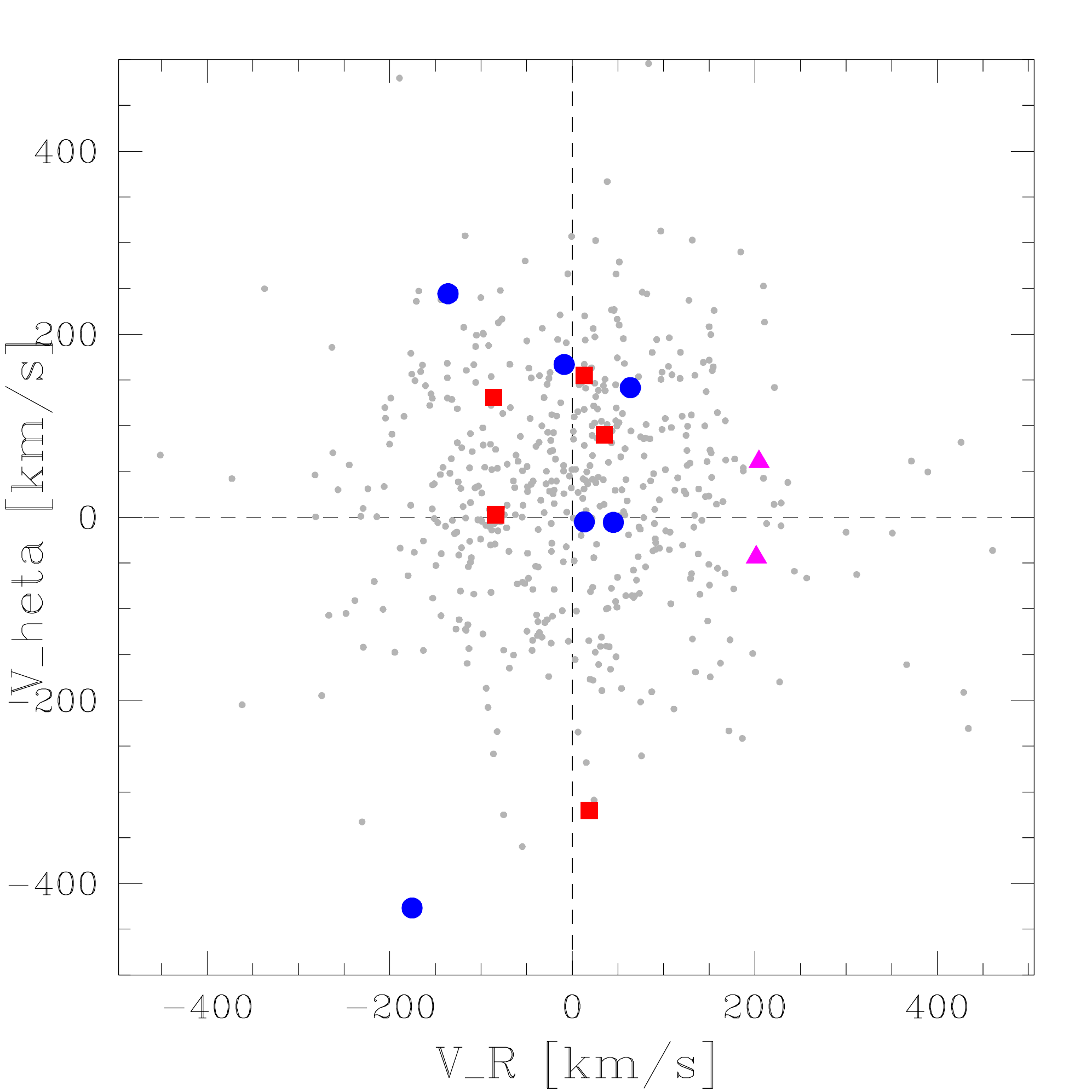}
\includegraphics[height=6.0cm]{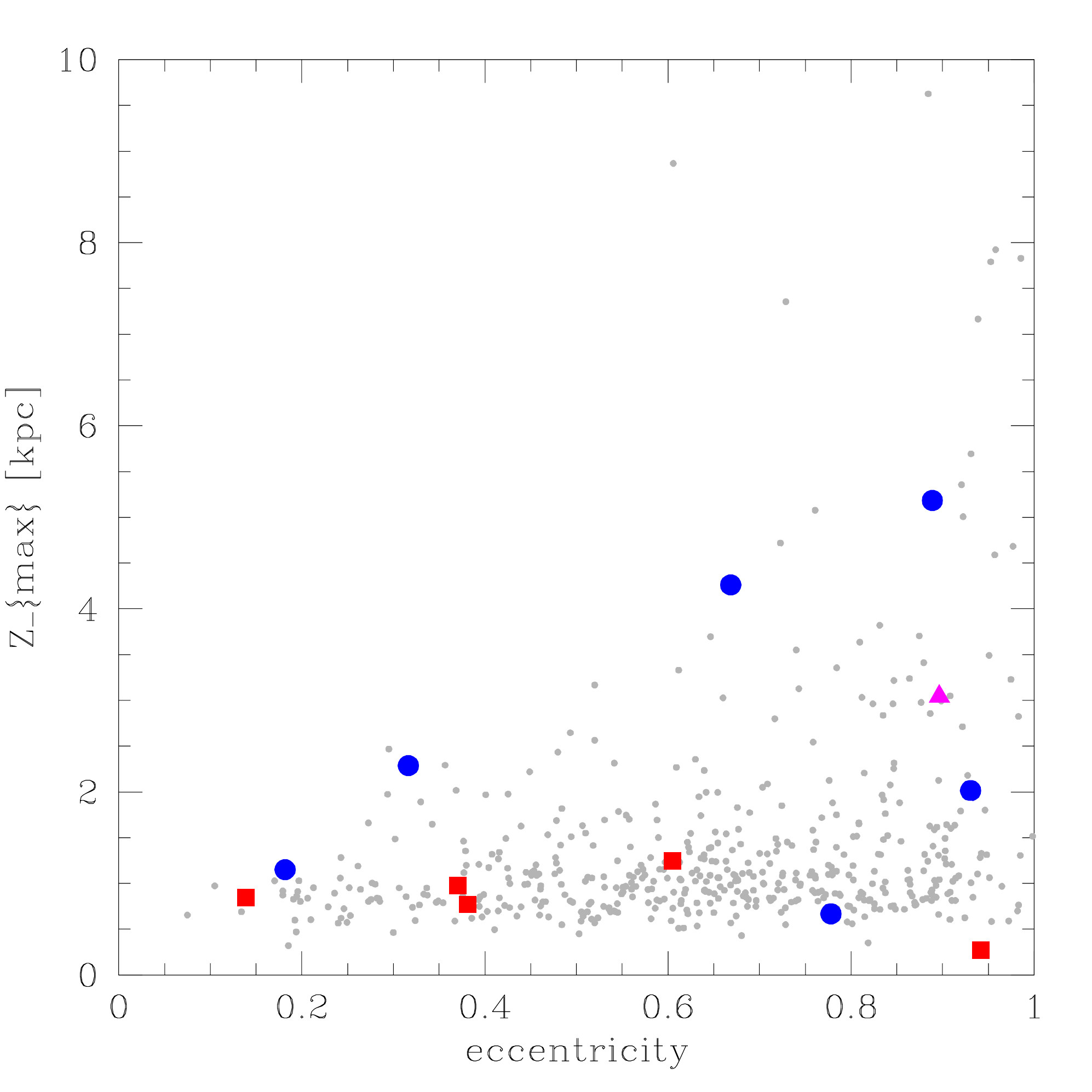}
\includegraphics[height=6.0cm]{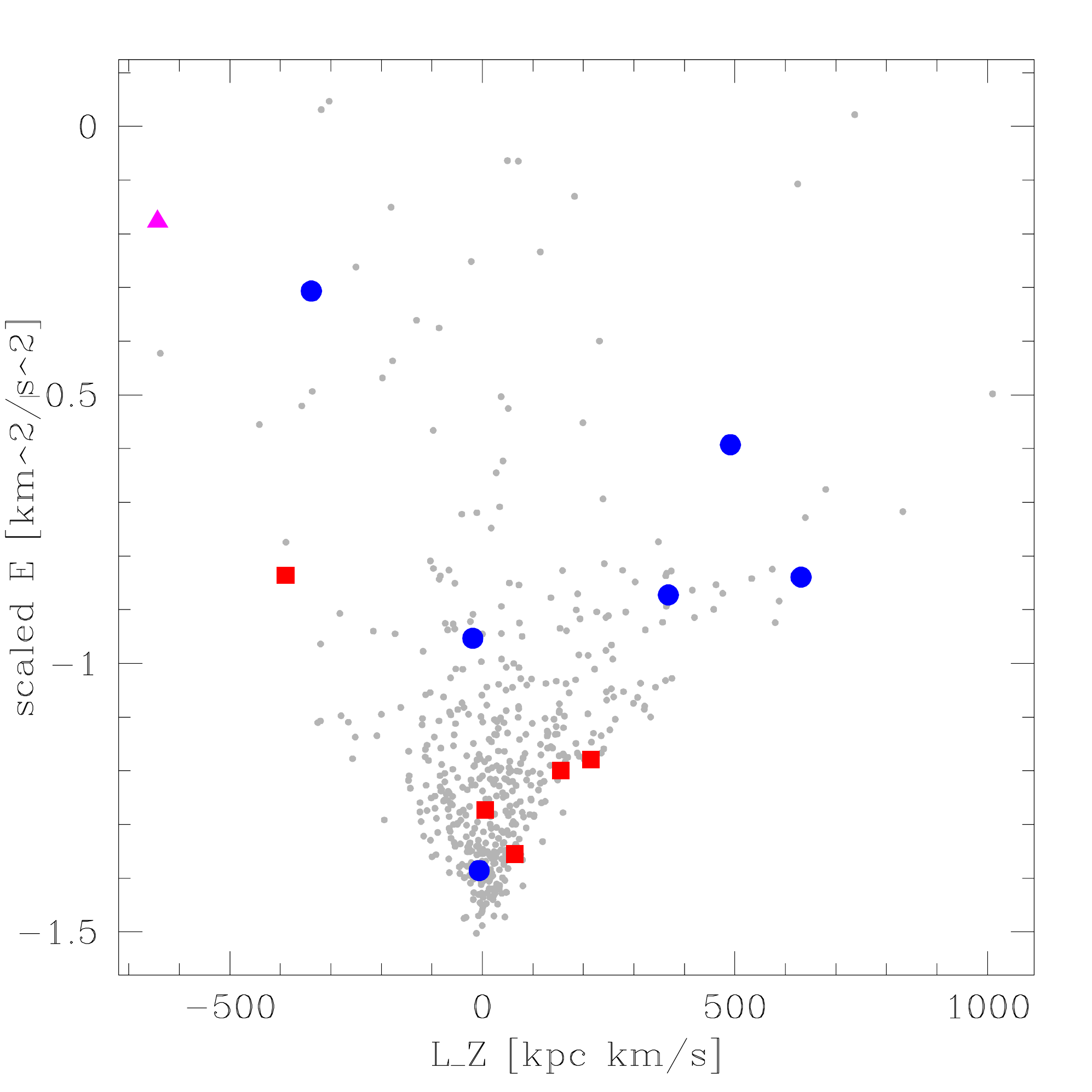}
}
\caption{{\it Left:} Behavior of the velocity components in spherical polar coordinates, namely 
radial $v_r$ and azimuthal $v_\theta$, for the RR01 stars presented here.  The symbols are the 
same as Figure~1.  The bulge RRLs from \citet{kunder16} are shown in grey.  
{\it Middle:} The eccentricity and $Z_{max}$ distance of our RR01 stars.
{\it Right:} The distribution of total Energy and $z$-angular momentum, 
$L_z$, where $L_z$ is the angular momentum 
out of the plane of the Galaxy's disk.  The total Energy is divided by $10^5$.}
\label{radial_params}
\end{figure*}

Figure~\ref{radial_params} shows the spherical polar components of the velocities (radial $\rm v_r$, and 
azimuthal/tangental $\rm v_\theta$) of the RR01 stars presented here over-plotted on the RRLs 
from \citet{kunder16}.  There is no obvious similarity in these velocities. 
The pulsationally clumped RR01s also show a wide range of eccentricities, $Z_{max}$ distances, 
angular momentum and energies.  The uncertainties on our orbital parameters 
are considerable -- in some cases, greater than 100\% due mainly to the 
distance and radial velocity uncertainties of the RR01 stars (see e.g., Figure~\ref{distcomp}).
However, the general trends of the distribution of the orbital parameters remain; 
the pulsationally clumped RR01s, even within their respective uncertainties, do not form
any discernible orbital similarities.

We cannot rule out that within the pulsationally clumped RR01s, a subset belong to a moving group.  
For example, 14029 and 15964 have radial velocities within 1 sigma of each other, and similar proper 
motions.  They are spatially $\sim$2 degrees apart from each other.  
A larger sample of RR01s with radial velocities, radial velocity estimates with smaller uncertainties, 
and detailed abundance information could give more insight on moving groups within 
the pulsationally clumped bulge RR01s.  

In general, the bulge RR01 stars have apocenter distances less than $\sim$4~kpc.  The exception is 
07686 (the RR01 star with a radial velocity of $\sim$441 km~s$^{-1}$) and the Sgr RR01 stars.
This indicates that most of the RR01 stars are confined to the inner Galaxy, but that roughly $\sim$15\%
are halo outliers just passing through the bulge\citep[see also][]{kunder15}.

\section{Conclusion}
The RR01 stars belonging to a compact group of RR01 stars in pulsation space, 
with period ratios of $P_{fo}/P_{f}\sim$0.74 and $P_{f}\sim$0.44, possess a large 
range of radial velocities and proper motions, and
are therefore not all moving together in a coherent structure.  
This is in contrast to the suggestion that these 28 stars are associated with a 
stellar stream that nearly vertically crosses the bulge, 
similar to the tidal stream of the Sgr Galaxy \citep{soszynski14}.  Spatial 
coherence would indicate that these stars were remnants of a relatively recent 
merger (like the Sgr merger), 
as the positional signature of the stars is still intact, and hence that some
kinematic signature would be retained.  

Although we find that the Sgr RR01 stars have similar radial velocities, 
proper motions, and period ratios, this is not the case 
for the pulsationally clumped RR01s, whose radial velocities span $\sim$300 km/s and whose 
proper motions span 12 mas~yr${-1}$.  Most of the bulge RR01 stars have orbits that 
confine them to the inner $\sim$4~kpc of the Galaxy, but we are not able to find any coherence 
in the pulsationally clumped RR01s and
3D velocities, angular momentum, or orbital parameters.
If the pulsationally clumped RR01s are indeed the relic of a stellar cluster or a dwarf galaxy disrupted by 
tidal interactions with the Milky Way, any kinematic signature has now been lost.  

\acknowledgements
We thank the Australian Astronomical Observatory, which have made these observations possible. 
The grant support provided, in part, by the M.J. Mudrock Charitable Trust (NS-2017321) is acknowledged.

\clearpage
\begin{table}
\begin{scriptsize}
\centering
\caption{Radial Velocities of double-mode RR Lyrae stars}\label{lcpars}
\begin{tabular}{lllllllllll} \hline OGLE ID & RA & Dec & $HRV_{\phi=.38}(km/s)$ & \# Epochs & $Period_f$(days) & $Period_{fo}$(days) & $(V)_{mag}$ & $(I)_{mag}$ & $I_{amp}$ & Dist(kpc) \\ \hline
\hline
09692$^c$* & 17:59:04.69 & -26:59:45.80 & 191 & 1 & 0.44175049 & 0.32690731 & 19.475 & 16.441 & 0.023 & 6.8 \\
14029$^c$ & 18:07:54.91 & -29:07:06.60 & 14 & 2 & 0.42994289 & 0.31798626 & 16.417 & 15.399 & 0.08 & 6.8 \\
14523$^c$ & 18:09:07.56 & -28:28:49.70 & -115 & 3 & 0.43271139 & 0.31999264 & 17.522 & 16.122 & 0.064 & 9.3 \\
10796$^c$ & 18:01:01.15 & -27:43:43.90 & 63 & 2 & 0.43137749 & 0.31910024 & 17.274 & 15.774 & 0.07 & 6.3 \\
15964$^c$ & 18:14:57.56 & -28:01:36.60 & 31 & 4 & 0.43885161 & 0.32474475 & 16.08 & 15.252 & 0.057 & 7.7 \\
16665$^c$ & 18:22:58.23 & -25:50:29.30 & 127 & 1 & 0.43427162 & 0.32112679 & 16.696 & 15.575 & 0.029 & 7.5 \\
11158 & 18:01:42.64 & -27:58:21.70 & -2 & 2 & 0.44178959 & 0.32624505 & 18.285 & 16.486 & 0.273 & 11.9 \\
07686 & 17:55:54.44 & -36:38:58.50 & 441 & 1 & 0.46652248 & 0.34684518 & 16.443 & 15.377 & 0.212 & 8.2 \\
10510 & 18:00:30.52 & -28:26:37.90 & -55 & 2 & 0.50307068 & 0.37496547 & 16.078 & 14.844 & 0.114 & 5.6 \\
14915 & 18:10:19.99 & -28:00:22.50 & -67 & 3 & 0.43338541 & 0.31699525 & 16.496 & 15.417 & 0.219 & 7.1 \\
16162 & 18:16:15.53 & -28:34:36.60 & -112 & 3 & 0.47774077 & 0.35538199 & 16.318 & 15.553 & 0.176 & 10.1 \\
35964 & 18:18:38.12 & -26:04:32.60 & -7 & 4 & 0.54309504 & 0.40511131 & 16.842 & 15.679 & 0.1 & 11.7 \\
16386* & 18:18:16.99 & -25:12:52.40 & -11 & 3 & 0.542311 & 0.40460239 & 17.296 & 16.304 & 0.135 & 13.6 \\
34653$^{sgr}$ & 18:12:08.67 & -28:57:31.80 & 169 & 1 & 0.47650339 & 0.35444355 & 18.966 & 18.02 & 0.192 & 29.7 \\
34704$^{sgr}$ & 18:12:22.91 & -28:49:34.80 & 178 & 1 & 0.47478192 & 0.35291588 & 18.297 & 17.364 & 0.094 & 23.0 \\
 \hline
\hline
\end{tabular}
\hspace*{-4.5cm} \ \ \ \ \ * no Gaia DR2 proper motion \\
\hspace*{-5.7cm}  \ \ \ \ \ $^c$ period ratio clump \\
\hspace*{-5.7cm}  $^{sgr}$ Sgr dwarf galaxy \\
\end{scriptsize}
  \end{table}

\end{document}